\documentstyle[a4,12pt,epsfig]{article}
\begin{document}

\hbox{}
\noindent  January 15, 2000  \hfill HU--EP--00/06

\vspace*{1.5cm}

\begin{center}

{\large
HALF-INTEGER TOPOLOGICAL CHARGES BELOW AND  \\ 
ABOVE THE DECONFINEMENT TRANSITION?
\footnote{Presented by A. I. Veselov at the NATO Advanced Research
Workshop {\it Lattice Fermions and Structure of the Vacuum},
October 1999, Dubna, Russia.}
}
\vspace*{0.8cm}

E.-M. Ilgenfritz$^a$, 
M. M\"uller--Preussker$^b$, 
and A. I. Veselov$^c$

\vspace*{0.3cm}

\end{center}

{\sl \noindent
\hspace*{15mm} $^a$  
University of Kanazawa, Institute of Theoretical Physics, \\ 
\hspace*{18mm}  Kanazawa 920-1192, Japan
  \\
\hspace*{15mm} $^b$ 
Humboldt-Universit\"at zu Berlin, Institut f\"ur Physik, \\
\hspace*{18mm}  D--10115 Berlin, Germany
  \\
\hspace*{15mm} $^c$ 
Institute of Theoretical and Experimental Physics, \\
\hspace*{18mm}  Moscow, 117259, Russia
 } 
\par\vspace{1.5cm}\noindent
{\bf Abstract.}
For pure $SU(2)$ lattice gauge theory at finite $T$, 
by the help of the cooling method, 
we search for classical (approximate) solutions having non-trivial 
holonomy at the spatial boundary. We identify various typical objects 
and provide their relative frequency of occurence for
the confinement and deconfinement phases. Among the configurations
obtained we see also the dissociated BPS monopole pairs recently discussed 
by van Baal and collaborators.

\vspace{1.0cm}
\section{Introduction}
During the last two years, the carriers of topological charge 
in a Yang-Mills field at finite temperature (calorons) have 
been thoroughly reconsidered by van Baal and co-workers. The
progress has been summarized at this conference \cite{this_conference}. 
It has been demonstrated that completely
different caloron solutions appear once a non-trivial holonomy 
${\cal P}({\bf x})$ at ${|\bf x}|\rightarrow\infty$ is admitted; 
the Polyakov line $L({\bf x})=\frac{1}{2}{\mathrm tr} {\cal P}({\bf x})$  
is the trace of the holonomy.
Prior to this development, semiclassical models at finite temperature were 
based exclusively on properties of periodic instantons, being classical 
solutions with trivial holonomy {\it i.e.} 
${\cal P}({\bf x}) \rightarrow 1$ for ${|\bf x}|\rightarrow\infty$
('t~Hooft periodic instanton)
\cite{t_hooft,harrington_shepard,gross_pisarski_yaffe}.

The outstanding feature of these {\it new calorons} is the fact that 
(within a certain parameter range) monopole constituents of an
instanton can become explicit as degrees of freedom
\cite{instanton_constituents}.  They carry magnetic as well as electric charge 
(in fact, they are Bogomol'ny-Prasad-Sommerfield (BPS) 
monopoles or `dyons') and $1/N_{\mathrm{color}}$
units of topological charge. Being part of classical solutions of 
the Euclidean field equations, one can hope that the instanton 
constituents can play an independent role in a semiclassical
analysis of $T\ne0$ Yang-Mills theory (and of full QCD). 

The variety of selfdual solutions for various topological charge $Q$ 
has been discussed as classical solutions on
the lattice with nontrivial holonomy from twisted boundary conditions 
\cite{caloron_lattice}.

The nontrivial holonomy {\it per se} is no obstacle to a semiclassical 
approach if that is not restricted to the one-instanton approximation. 
To what extent such a semiclassical description is reliable (and 
exhaustive), has to be investigated for each phase (confinement 
and deconfinement) of pure Yang-Mills theory.

In exploratory studies we have searched for characteristic differences 
between the two phases as far as quasiclassical background fields are
concerned. These become visible in the result of cooling. 

In a previous study \cite{we} we discussed finite temperature 
$SU(2)$ lattice gauge theory in a finite spatial box with specific 
boundary conditions. The latter were chosen such that the finite
(not too large) system was put into a definite single-monopole background
field. By varying the monopole scale we could study both the situations: the
purely magnetic 't Hooft-Polyakov (HP) monopole and the self-dual BPS monopole
(`dyon'). We observed a specific influence of these different boundaries
on the quantum state inside the box. Whereas the HP monopole has turned out to 
favour deconfinement, the BPS monopole has been keeping the system
in the confinement even at temperatures larger then the usual critical one. 
Let us note that these boundary conditions are characterized by  
different holonomy values, too. 

In comparison with the previous study in our present work 
we have employed simpler spatial boundary conditions.
We have fixed and have left untouched under cooling only the boundary 
time-like link variables in order to keep a certain value of 
${\cal P}({\bf x})={\cal P}_{\infty}$ everywhere on the spatial 
surface of the system while conserving periodicity. 

In this study, the influence of the respective phase, that we want to
describe, is twofold : ({\it i}) the cooling starts from genuine thermal
Monte Carlo gauge field configurations, generated on a $N_s^3\times N_t$
lattice; ({\it ii}) the value of the holonomy ${\cal P}_{\infty}$ was 
chosen in accordance with the average of $L$, which is vanishing in 
the confinement phase and nonvanishing in deconfinement.

\section{Results}
We consider $SU(2)$ lattice gauge theory with the standard Wilson action.
Our cooled samples are obtained from Monte Carlo ensembles 
on a $16^3\times 4$ lattice. For $N_t=4$, the gauge coupling $\beta=2.2$ 
stands for the confinement phase with 
$\langle L \rangle \simeq 0.$
and $\beta=2.4$ for the deconfinement phase with 
$\langle L \rangle =0.27$, respectively.
The timelike links $U_{x,\mu=4}$ are frozen at the spatial boundary
equal to each other such that $(U_{x,\mu=4})^{N_t}={\cal P}_{\infty}$.  
For the holonomy itself, an `Abelian' form 
${\cal P}_{\infty}=a_0 + i~a_3~\tau_3$ was chosen, with
$a_0=~\langle L \rangle$ and $a_3=\sqrt{1-a_0^2}$ in correspondence with
the average Polyakov line. The cooling method chosen was the simplest, 
rapid relaxation method keeping the Wilson action. 
In order to search exclusively for objects with low action  
the criterion for the first stopping at some 
cooling step $n$ was that $S_n < 2~S_{\mathrm{inst}}$, the last change 
of action $|S_n - S_{n-1}| < 0.01~S_{\mathrm{inst}}$, and that  
the second derivative $S_n-2~S_{n-1}+S_{n-2} < 0$ ($S_{\mathrm{inst}}$ 
denoting the action of a single instanton). 
For each $\beta$-value we have scanned $O(200)$ configurations obtained
by cooling.

The cooled sample taken from the confinement phase has a clearly 
different composition compared to that from the deconfinement ensemble. 
This is listed in Table 1 which shows the relative occurrence
of the few characteristic types of nonperturbative configurations.
In the following let us explain these configurations in some detail.
\begin{table}[h]
\caption{\it Relative frequencies of the occurence of different kinds of
solutions as explained in the text for $\beta=2.2$ (confinement) and
$\beta=2.4$ (deconfinement) after cooling.}
\begin{center}
\begin{tabular}{lcc}
\hline
Type of solution     & $ \beta=2.2 $        &  $ \beta=2.4 $        \\
\hline
$DD$                 & $ 0.63  \pm 0.08  $  &  $ 0.02  \pm 0.01  $  \\
$D\overline{D}$      & $ 0.27  \pm 0.05  $  &  $ 0.78  \pm 0.07  $  \\
$CAL$                & $ 0.02  \pm 0.01  $  &  $ 0.              $  \\
$M$, $2M$            & $ 0.01  \pm 0.01  $  &  $ 0.07  \pm 0.02  $  \\
trivial vacuum       & $ 0.07  \pm 0.03  $  &  $ 0.13  \pm 0.03  $  \\
\hline
\end{tabular}
\end{center}
\end{table}
\par\medskip\noindent
{\it Confinement phase.}  
Here dominate selfdual (or antiselfdual) configurations, 
and among them `dyon-dyon' pairs ($DD$) which are reminiscent of
the {\it new caloron} solutions.  In Fig. 1 we show, projected onto the 
$x_1-x_2$-plane ({\it i.e.} summed over $x_3,x_4$ or $x_3$, resp.), 
the topological charge 
and the Polyakov line of such a `dyon' pair. Notice the opposite sign of 
the Polyakov line near to the two same-sign topological charge bumps.
\begin{figure}
 \begin{minipage}{12.5cm}
 \begin{center}
  \epsfig{file=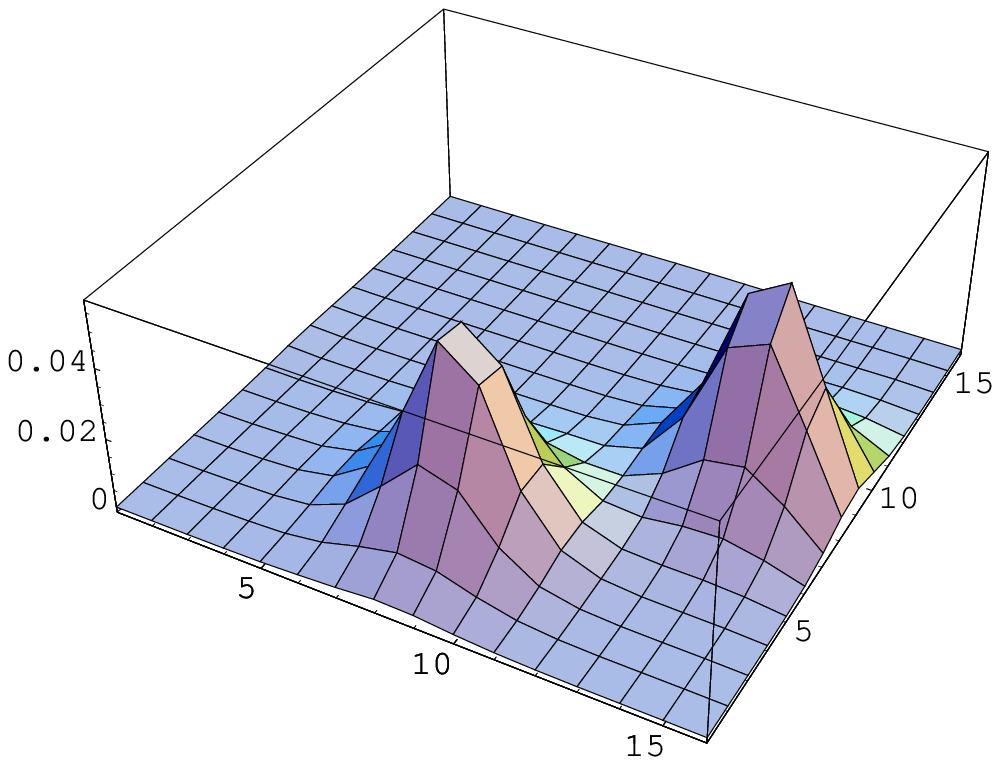,width=5.8cm,height=5.0cm} \hspace{5mm}
  \epsfig{file=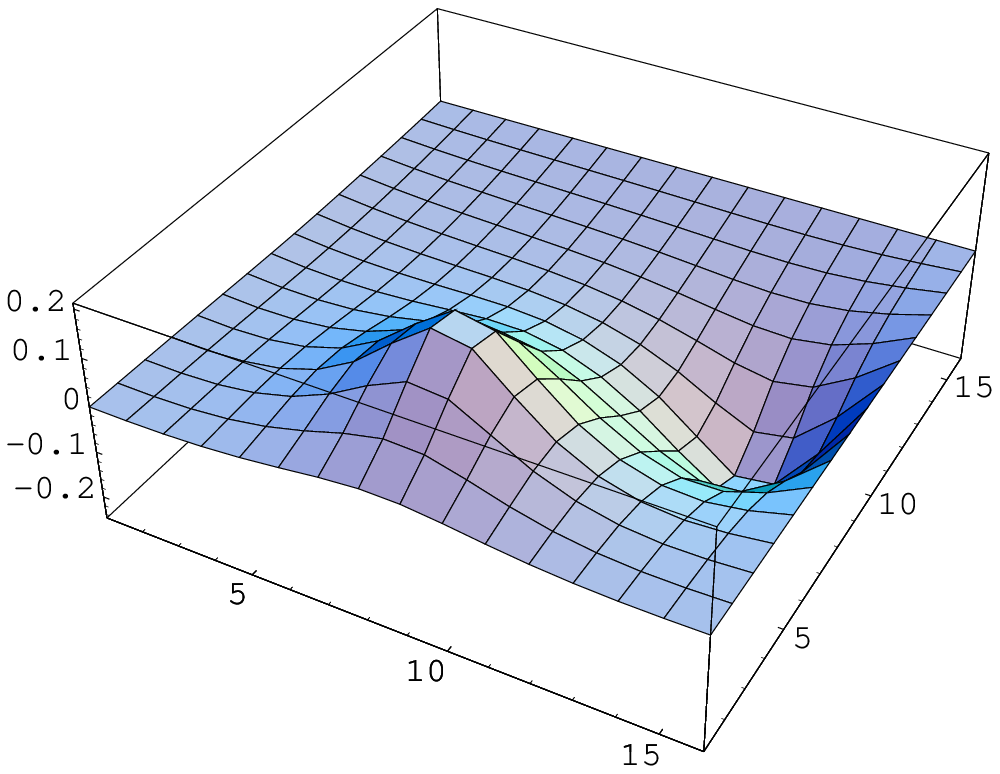,width=5.8cm,height=5.0cm}\\
  \vspace{-2mm}
(a) \hspace{5.5cm} (b)\\
 \end{center}
 \end{minipage}
 \caption{\it
(a) The $2D$ projected distribution of topological charge for a
selfdual `dyon-dyon' pair ($DD$) discovered by cooling in a $16^3\times 4$ 
thermal configuration generated at $\beta=2.2$;
(b) Similarly for the Polyakov line.}
\end{figure}

Other selfdual objects, having a rather $O(4)$ rotationally invariant 
distribution of action and topological charge, are frozen out relatively 
infrequently. They resemble the 't Hooft periodic instanton. We call them 
caloron ($CAL$), shown in Fig. 2 . Under the specific boundary
conditions, however, the structure of the Polyakov line around the caloron
is nontrivial in the sense that it has the opposite peaks of the Polyakov
line near the center of the action (and topological charge) distribution.
Thus, this type of configurations appears as a limiting case of the 
`dyon-dyon' pairs.
\begin{figure}
 \begin{minipage}{12.5cm}
 \begin{center}
  \epsfig{file=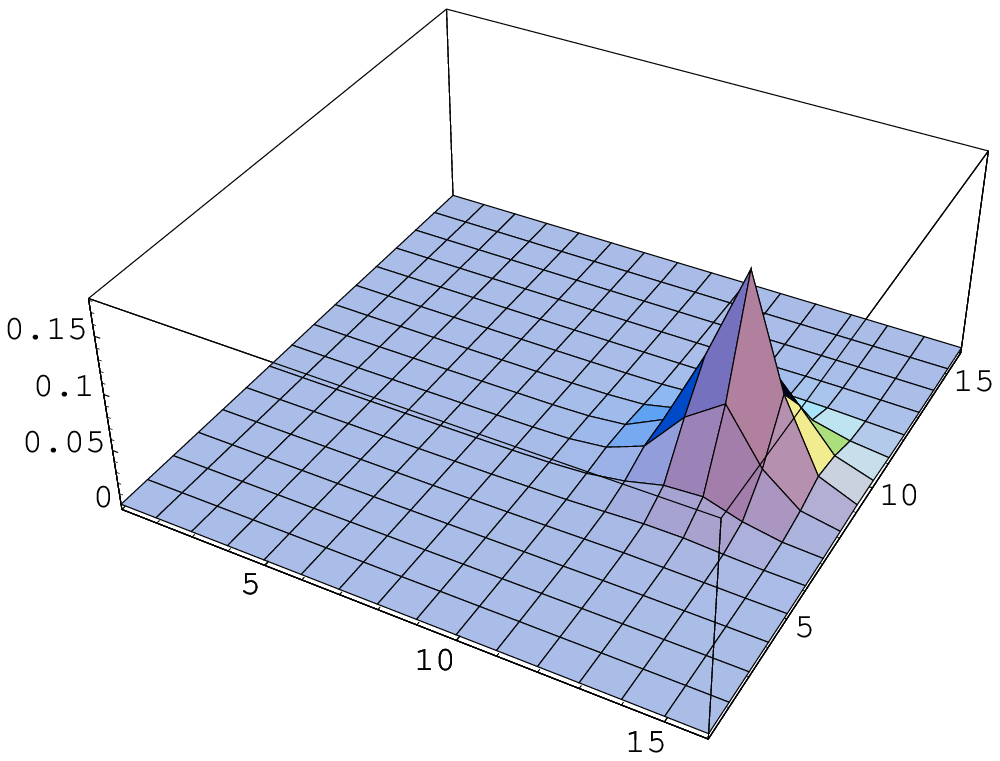,width=5.8cm,height=5.0cm} \hspace{5mm}
  \epsfig{file=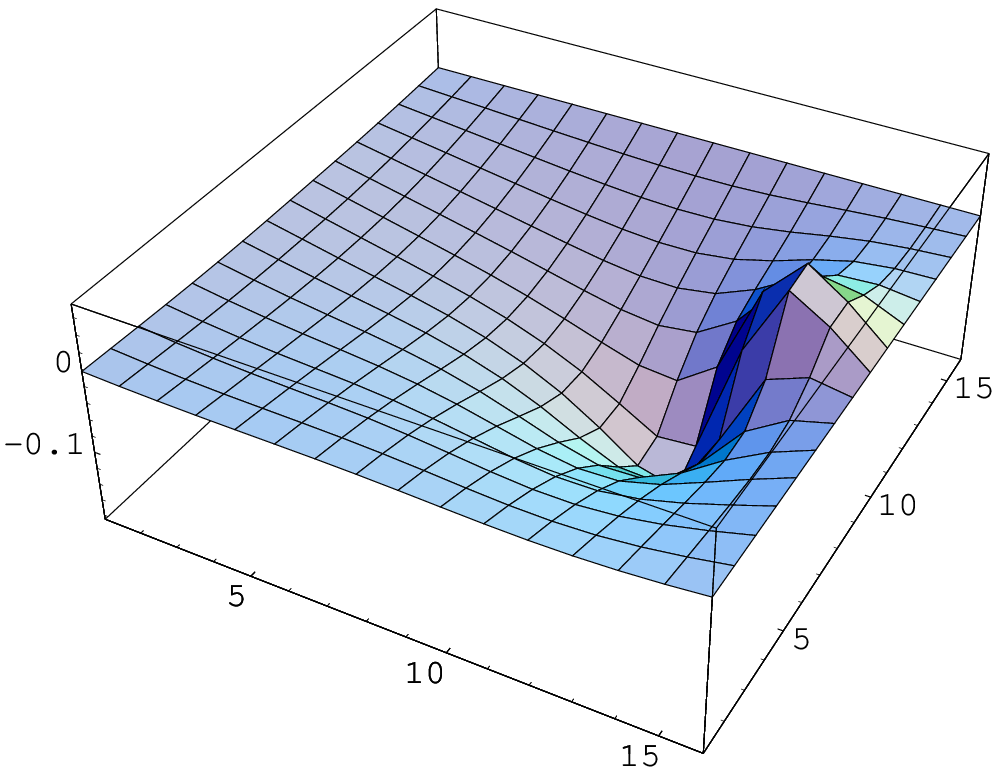,width=5.8cm,height=5.0cm}\\
  \vspace{-2mm}
(a) \hspace{5.5cm} (b)\\
 \end{center}
 \end{minipage}
 \caption{\it
(a) The $2D$ projected distribution of topological charge for a
selfdual, rotationally symmetric caloron ($CAL$) discovered by cooling in a 
$16^3\times 4$ thermal configuration generated at $\beta=2.2$;
(b) Similarly for the Polyakov line.}
\end{figure}

Mixed configurations with two lumps of opposite topological charge 
are found in a quarter of the configurations. We call them `dyon-antidyon' 
pairs ($D\overline{D}$). In fact, they are typical for the deconfined 
phase and below we discuss an example taken from $\beta=2.4$ . 
In these configurations the Polyakov line has a same-sign maximum on 
top of the opposite-sign topological charge lumps. Besides of this, 
the two sums $Q_{+}=\sum_x q(x)~\Theta(q(x))$ and
$Q_{-}=\sum_x q(x)~\Theta(-q(x))$ are almost equal to $+\frac12$ and
$-\frac12$, respectively, which supports an interpretation as 
half-instanton and half-antiinstanton. 

When one allows the holonomy ${\cal P}({\bf x})$ freely to relax as in
usual cooling with periodic boundary conditions, configurations like 
$DD$ and $D\overline{D}$ do not survive.

\par\medskip\noindent
{\it Deconfinement phase.}  
It is remarkable that selfdual or antiselfdual $DD$ configurations
are very rare in this case. $D\overline{D}$ mixed configurations are 
typical for the deconfined phase. For one example we show in Fig. 3 
the topological charge and Polyakov line (similar to Fig. 1). 
\begin{figure}
 \begin{minipage}{12.5cm}
 \begin{center}
  \epsfig{file=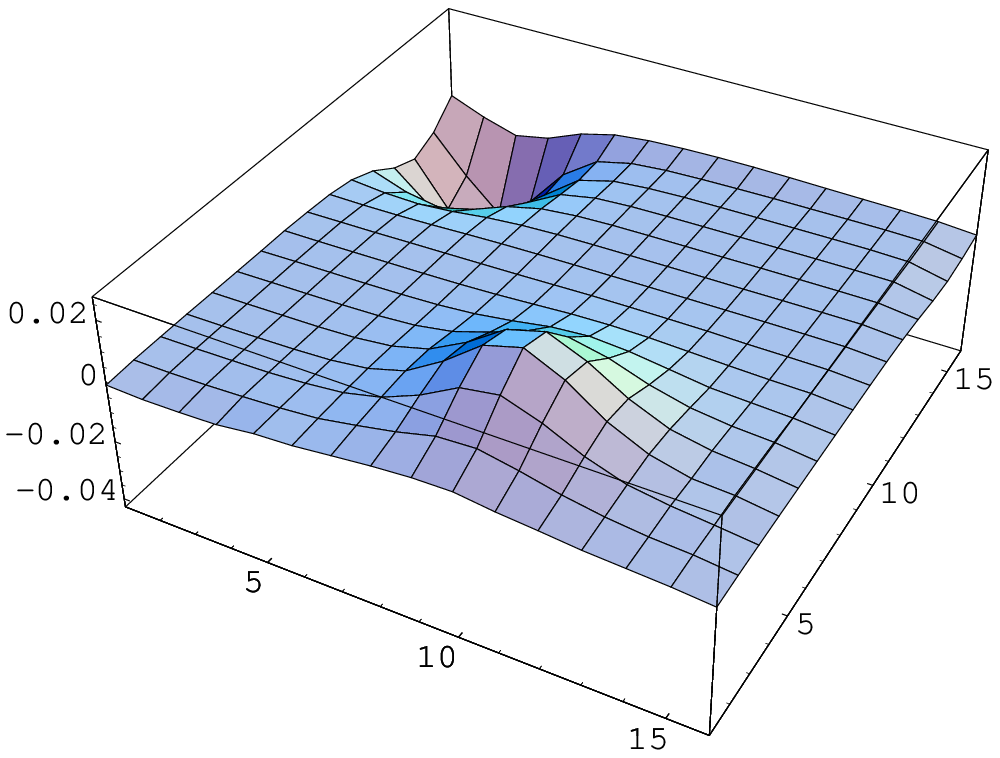,width=5.8cm,height=5.0cm} \hspace{5mm}
  \epsfig{file=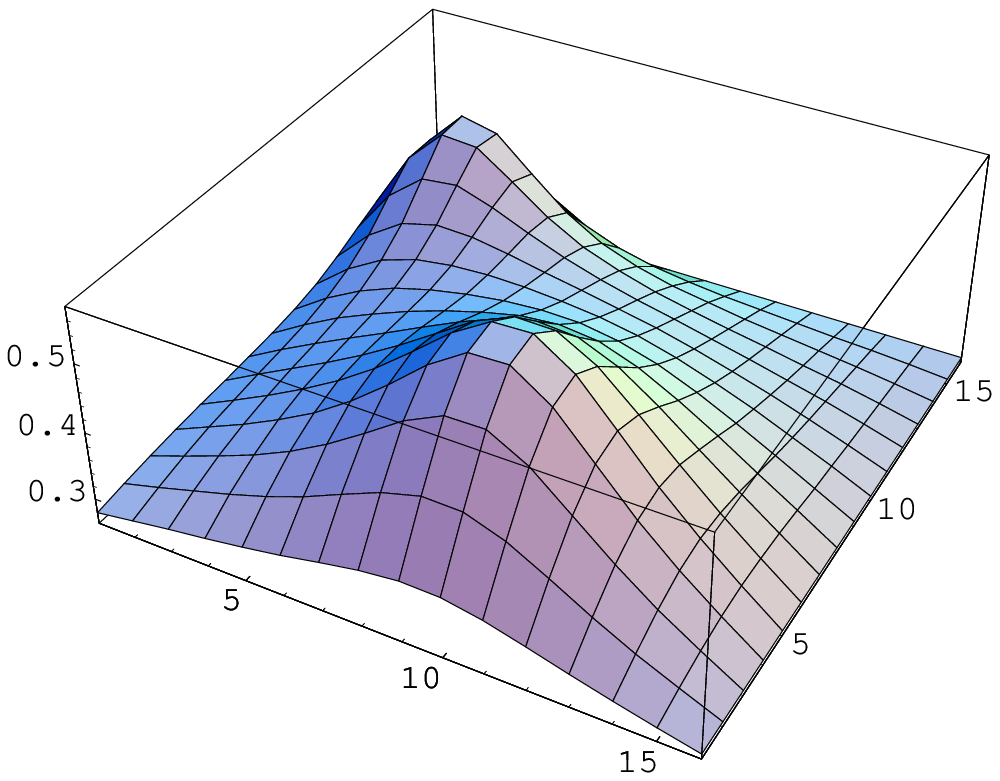,width=5.8cm,height=5.0cm}\\
  \vspace{-2mm}
(a) \hspace{5.5cm} (b)\\
 \end{center}
 \end{minipage}
 \caption{\it
(a) The $2D$ projected distribution of topological charge for a
mixed `dyon-antidyon' pair ($D\overline{D}$) 
discovered by cooling in a $16^3\times 4$ 
thermal configuration generated at $\beta=2.4$;
(b) Similarly for the Polyakov line.}
\end{figure}

In the deconfined phase the next important type of cooled configurations
are purely magnetic ones 
($S_{\mathrm{magnetic}}>>S_{\mathrm{electric}}$) with 
quantized action in units of $S_{\mathrm{inst}}/2$. We call these $M$ 
configurations.
With a smaller probability also magnetic configurations with twice as
large action ($2M$ type configurations) are found.
In all projections, the action is constantly distributed over the 
lattice with a high precision. A closer look at the different field 
strength components reveals that the action of $M$ type 
configurations resides only in a single magnetic field strength component
(in one $3D$ direction), while in the case of $2M$ type configurations 
two such fluxes, generically orthogonal to each other, are present.  
After fixing the maximally Abelian gauge (see below) 
these configurations turn out 
to be completely Abelian. Therefore, we can identify these configurations
as pure equally distributed magnetic fluxes which should be related
to world-sheets of Dirac strings (`Dirac sheet') on the dual lattice.
With some rate they also emerge in the result of further 
cooling of $D\overline{D}$ configurations. 
Such configurations are present in the confinement phase as well, 
but occur with tiny probability only.

Maximizing with respect to gauge transformations the gauge functional 
$R=\sum_{x\mu}{\mathrm tr}~\left(\tau_3~U_{x\mu}~\tau_3~U_{x\mu}^+\right)$,
we have put all these configurations into the maximally Abelian gauge
and have measured their `Abelianicity', {\it i.e.} $R_{\mathrm{max}}$
per link.
The $DD$ and $D\overline{D}$ configurations have an 
Abelianicity of 99.8\% (independent of the phase where they are found)
The rotationally invariant caloron is somewhat more Abelian
(99.9\%). But for the purely magnetic configurations we found an Abelianicity
of exactly 100\%.

Employing the Abelian projection, 
$U(1)$ mo\-no\-po\-les \cite{degrand_toussaint}
can be localized. 
The $DD$ and $D\overline{D}$ configurations are found to be static
to a high precision, and
Abelian monopole worldlines were observed to coincide with the `dyon' or 
`antidyon' position, again irrespective of the phase where these background
configurations have been extracted.
The non-static `caloron' configuration, however, is typically enclosed 
by a small Abelian monopole loop (of length 6).  

\section{Conclusions}
We have studied finite temperature Yang-Mills lattice fields 
with given non-trivial holonomy at the spatial boundary of a finite box.
Starting from Monte Carlo equilibrium configurations by cooling we
have found quasi-stable solutions in accordance with that (periodic) boundary
condition. The ensembles of solutions obtained strongly depend upon whether
we are in the confinement or in the deconfinement phase.
Most typically we observe `dyon-dyon' pairs within the confinement phase,
i.e. selfdual solutions of the type discussed by van Baal and collaborators
\cite{this_conference}. However, in the deconfinement
phase  `dyon-antidyon' solutions dominate. 
The latter objects have to be understood analytically. We did not find
pure magnetic HP-like monopoles as seen in \cite{laursen_schierholz}, where
cooling had been applied to finite temperature fields with purely
periodic boundary conditions. The latter monopoles - always accompanied
by a spurious opposite charged monopole - have trivial holonomy and, 
thus, could not show up in the present analysis.

We feel that the development of a semiclassical approach based on
solutions with non-trivial holonomy, i.e. in a mean-field-like setting, 
might have a chance to shed light on the mechanisms of the deconfinement 
transition.

\section*{Acknowledgements}
The authors are grateful to P.~van~Baal, B.~V.~Martemyanov, S.~V.~Molodtsov,
M.~I.~Polikarpov, A.~van~der~Sijs, Yu.~A.~Simonov, and J.~Smit 
for useful discussions.
This work was partly supported by RFBR grants N~97-02-17491
and N~99-01-01230 as well as by the joint RFBR-DFG project grant 
436~RUS~113/309~(R) and the INTAS grant 96-370.


\newpage
\begin{thebibliography}{99}

\bibitem{this_conference}
P. van Baal (1999), these Proceedings, e-Print Archive: hep-th/9912035. 

\bibitem{t_hooft}
G. 't Hooft (1976), unpublished; \\ see R. Jackiw, C. Nohl, and C. Rebbi
(1977), {\it Phys. Rev.}, {\bf D15}, p. 1642.

\bibitem{harrington_shepard}
B.J. Harrington and H.K. Shepard (1978), {\it Phys. Rev.}, {\bf D17}, p. 2122;
(1978), {\it Phys. Rev.}, {\bf D18}, p. 2990.

\bibitem{gross_pisarski_yaffe}
D.J. Gross, R.D. Pisarski, and L.G. Yaffe (1983), {\it Rev. Mod. Phys.}, 
{\bf 53}, p. 43.

\bibitem{instanton_constituents} 
T.C. Kraan and P. van Baal (1998), {\it Phys. Lett.}, {\bf B435}, p. 389. 

\bibitem{caloron_lattice}
M. Garcia Perez, A. Gonzalez-Arroyo, A. Montero, and P. van Baal (1999), 
{\it JHEP} {\bf 06}, p. 001; e-Print Archive: hep-lat/9903022. 

\bibitem{we}
E.-M. Ilgenfritz, S.V. Molodtsov, M. M\"uller-Preussker, and
A.I. Veselov (1999), {\it Eur. Phys. J.}, {\bf C8}, p. 335. 

\bibitem{degrand_toussaint} 
T.A. DeGrand and D. Toussaint (1980), {\it Phys. Rev.}, {\bf D22}, p. 2478. 

\bibitem{laursen_schierholz}
M.L. Laursen and G. Schierholz (1988), {\it Z. Phys.}, {\bf C38}, p. 501.  

\end{thebibliography}
\end{document}